\documentclass[%
reprint,
superscriptaddress,
showpacs,
preprintnumbers,
 bibnotes,
 amsmath,amssymb,
 aps,
prl
]{revtex4-1}

\usepackage{graphicx}
\usepackage{dcolumn}
\usepackage{bm}
\usepackage{color}

\begin{document}


\title{Low-energy spin excitations in (Li$_{0.8}$Fe$_{0.2}$)ODFeSe superconductor studied with inelastic neutron scattering}

\author{Mingwei~Ma}
\thanks{These authors contributed equally to this study.}
\affiliation{International Center for Quantum Materials, School of Physics, Peking University, Beijing 100871, China}
\author{Lichen~Wang}
\thanks{These authors contributed equally to this study.}
\affiliation{International Center for Quantum Materials, School of Physics, Peking University, Beijing 100871, China}
\author{Philippe~Bourges}
\affiliation{Laboratoire L\'{e}on Brillouin, CEA-CNRS, Universit\'{e} Paris-Saclay, CEA Saclay, 91191 Gif-sur-Yvette, France}
\author{Yvan~Sidis}
\affiliation{Laboratoire L\'{e}on Brillouin, CEA-CNRS, Universit\'{e} Paris-Saclay, CEA Saclay, 91191 Gif-sur-Yvette, France}
\author{Sergey~Danilkin}
\affiliation{Bragg Institute, Australian Nuclear Science and Technology Organization, New Illawarra Road, Lucas Heights NSW-2234, Australia}
\author{Yuan~Li}
\email[]{yuan.li@pku.edu.cn}
\affiliation{International Center for Quantum Materials, School of Physics, Peking University, Beijing 100871, China}
\affiliation{Collaborative Innovation Center of Quantum Matter, Beijing 100871, China}

\begin{abstract}
We report an inelastic neutron scattering study of single crystals of (Li$_{0.8}$Fe$_{0.2}$)ODFeSe.  Temperature-dependent low-energy spin excitations are observed near $\mathbf{Q}=$ (0.5, 0.27, 0.5) and equivalent wave vectors symmetrically surrounding $\mathbf{Q}=$ (0.5, 0.5, 0.5) in the 1-Fe Brillouin zone, consistent with a Fermi-surface-nesting description. The excitations are broadly distributed in energy, ranging from 16 to 35 meV.  Upon cooling below the superconducting critical temperature ($T_\mathrm{c}$), magnetic response below twice the superconducting gap $2\Delta_\mathrm{SC}$ exhibits an abrupt enhancement, consistent with the notion of spin resonance, whereas the response at higher energies increases more gradually with only a weak anomaly at $T_\mathrm{c}$. Our results suggest that (Li$_{0.8}$Fe$_{0.2}$)ODFeSe might be on the verge of a crossover between different Cooper-pairing channels with distinct symmetries.

\end{abstract}

\pacs{74.70.Xa, 
78.70.Nx, 
74.20.Rp, 
74.25.Ha  
}

\maketitle
A pivotal issue concerning the Cooper-pairing mechanism in the Fe-based superconductors (FeSCs) is the pairing symmetry, which is commonly regarded as an important thread for distinguishing among different theoretical proposals. Pairing mediated by electron-phonon interactions \cite{BoeriPRL2008} and/or orbital fluctuations \cite{KontaniPRL2010} is expected to occur in the plain $s$-wave ($s^{++}$) channel, whereas pairing mediated by spin fluctuations \cite{ScalapinoRMP2012} is expected to have sign-reversal behaviors in the gap function and occur in the extended $s$-wave ($s^{+-}$) or the $d$-wave channel \cite{MazinPRL2008,KurokiPRL2008,SeoPRL2008,SiPRL2008}. When hole pockets are present at the $\Gamma$ point, it has been reasonably well established that the predominant pairing symmetry is $s^{+-}$ \cite{HanaguriScience2010,DaiRevModPhys2015}, which favors the unconventional pairing mechanism associated with spin fluctuations. But because the FeSCs are multi-orbital systems with multiple magnetic exchange interactions that are comparable in strength \cite{YildirimPRL2008}, the preference on one pairing channel over another may further depend on the Fermi-surface (FS) topology \cite{KurokiPRL2008,SaitoPRB2011}. Therefore, the assumption that a universal pairing mechanism applies to all FeSCs requires a more stringent test, namely, pairing symmetry needs to be determined and compared with theory for systems with very different band filling and FS topologies.

An important case of distinct FS topology was first established in alkali-metal intercalated Fe selenides \cite{GuoPhysRevB2010}, which have no hole pocket at the $\Gamma$ point as a result of heavy electron doping \cite{MouPRL2011,ZhangNatMater2011,QianPRL2011}. Later it became clear that the electron doping can be achieved with various methods, including intercalation into bulk FeSe \cite{YingSciRep2012,GuoNatureCommun2014,LuNatureMater2015,DongJACS2015,ZhaoLNatureCommun2015,NiuPhysRevB2015}, epitaxial growth of a single atomic layer of FeSe on a charge-transferring SrTiO$_3$ substrate (FeSe/STO) \cite{WangCPL2012,HeSLNatureMater2013,TanNatMater2013}, and surface dosing of FeSe with potassium \cite{MiyataNatMater2015,YePreprint2015,WenNatCommun2016}, all of which end up with similar electronic structures. It is intriguing that the absence of hole pockets at the $\Gamma$ point is empirically linked to the much higher values of $T_\mathrm{c}$ than in bulk FeSe \cite{HsuPNAS2008}. The associated pairing symmetry is considered to be of great theoretical importance but has remained unsettled \cite{SaitoPRB2011,FangPRX2011,MaierTAPhysRevB2011,WangEPL2011,MazinPRB2011,KhodasPRL2012,GuterdingPRB2015,NicaPreprint2015}, in part because of seemingly contradictory results from inelastic neutron scattering (INS) and scanning tunneling spectroscopy (STS) experiments: On the one hand, the observation of a spin resonance below $T_\mathrm{c}$ in INS studies of $A_x$Fe$_{2-y}$Se$_2$ ($A=$ Rb, Cs, K) \cite{ParkPhysRevLett2011,TaylarPhysRevB2012,FriemelEPL2012} points towards a gap function with opposite signs on the electron pockets \cite{MaierTAPhysRevB2011,KhodasPRL2012}. On the other hand, the inability of non-magnetic impurities to create in-gap states in FeSe/STO suggests that the pairing symmetry is plain $s$-wave \cite{FanNatPhys2015}. Notably, both of these systems have material-specific aspects that are not representative of all FeSCs. The superconductivity in $A_x$Fe$_{2-y}$Se$_2$ often suffers from a low volume fraction \cite{WangZPhysRevB2011}, probably due to the requirement of a high density of Se vacancies and/or of a particular type of interfaces \cite{LiPRL2012}, whereas the superconductivity in FeSe/STO may be strongly assisted by phonons in the substrate \cite{LeeNature2014,LiSciBull2016}.

In order to assess the universality of the previous results and address the contradictions, it is desirable to investigate a phase-pure bulk material without a substrate. The newly discovered (Li$_{0.8}$Fe$_{0.2}$)OHFeSe superconductor offers such an opportunity \cite{LuNatureMater2015,DongJACS2015}, since it does not show substantial chemical phase separation and is stable in air. Photoemission experiments have demonstrated that (Li$_{0.8}$Fe$_{0.2}$)OHFeSe has a FS topology representative of that of heavily-electron-doped FeSe sheets \cite{ZhaoLNatureCommun2015,NiuPhysRevB2015}. Here we report our INS study of fully-deuterated (Li$_{0.8}$Fe$_{0.2}$)ODFeSe single crystals, aiming to characterize the low-energy spin fluctuations. We find that the temperature-dependent magnetic signals are centered at momentum positions consistent with a FS-nesting picture, but the signal is rather widely distributed in energy and extends well above twice the superconducting energy gap $2\Delta_\mathrm{SC}$. While the enhancement of signals below $2\Delta_\mathrm{SC}$ has a temperature dependence consistent with that of a spin resonance in unconventional superconductors, the signal above $2\Delta_\mathrm{SC}$ exhibits a more gradual increase below $T_\mathrm{c}$. The fact that these two parts of signals are comparable in strength despite their distinct temperature dependence suggests that (Li$_{0.8}$Fe$_{0.2}$)OHFeSe, and perhaps heavily-electron-doped FeSe sheets in general, might host two types of pairing interactions that are favorable for distinct pairing symmetries.

Our triple-axis INS experiments were performed on the spectrometer 2T at the Laboratoire L\'{e}on Brillouin (LLB), France and on the spectrometer TAIPAN at the Bragg Institute of Australian Nuclear Science and Technology Organization. The sample consisted of over one hundred single crystals of fully-deuterated (Li$_{0.8}$Fe$_{0.2}$)ODFeSe, which were grown with a hydrothermal reaction method \cite{DongPhysRevB2015} and coaligned in the ($H$, $K$, $H$) scattering plane. Here and throughout the paper, reciprocal-space vectors are quoted in reciprocal-lattice units (r.l.u.) under the notation of the 1-Fe Brillouin zone, with unit-cell parameters $a = b = 2.65 \mathrm{\AA}$ and $c = 9.30 \mathrm{\AA}$. Clear $x$-ray Laue reflections and a sharp increase of diamagnetic signals below $T_\mathrm{c} = 39$ K demonstrate the high quality of our sample (Fig.~\ref{Fig1}). The INS data were collected in fixed-$k_\mathrm{f}$ mode ($k_\mathrm{f} = 3.85 \mathrm{\AA}^{-1}$) using a focusing pyrolytic graphite (PG) monochromator and analyzer. Additional PG filters were placed between the sample and the analyzer to eliminate higher-order contaminations.

\begin{figure}
\includegraphics[width=3in]{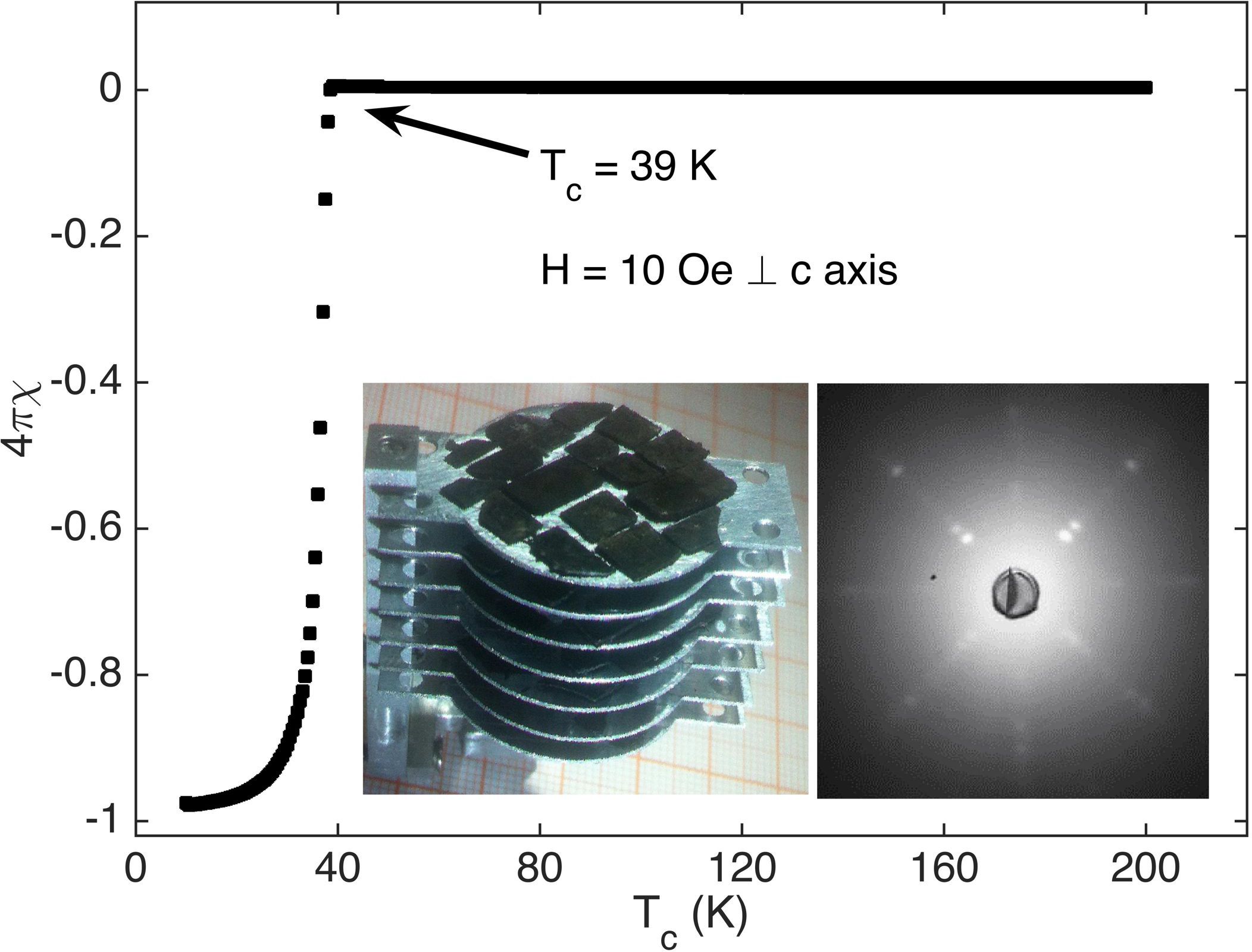}
\caption{\label{Fig1}
Temperature dependence of magnetic susceptibility of (Li$_{0.8}$Fe$_{0.2}$)ODFeSe measured with a magnetic field of 10 Oe applied perpendicular to the $c$ axis after zero-field cooling. Left inset: photo of our co-aligned sample for INS experiments. Right inset: $x$-ray Laue-reflection pattern from a single crystal ($x$-rays incident along the $c$ axis).
}
\label{Fig1}
\end{figure}

Figure~\ref{Fig2}(a) displays the result of momentum ($\mathbf{Q}$) scans at a fixed energy of 18 meV, performed along the trajectory (0.5, $K$, 0.5) and at temperatures both below and above $T_\mathrm{c}$. As an effort to search for the spin resonance, this trajectory was chosen after the previously reported results for Rb$_x$Fe$_{2-y}$Se$_2$ \cite{ParkPhysRevLett2011}, since the two materials have similar electronic structures. The energy was chosen to be about $64\%$ \cite{YuNatPhys2009} of the maximal superconducting gap $2\Delta_\mathrm{SC}$, which has been previously determined to be about 28.6 meV \cite{DuNatureCommun2016}. Despite the use of a fully-deuterated sample, the background scattering is very strong due to the sizable neutron incoherent scattering cross sections of the D ($^2$H) and Li isotopes. Therefore, we use the intensity difference between the two temperatures to extract the magnetic signals [Fig.~\ref{Fig2}(b)], and find a net intensity increase below $T_\mathrm{c}$ near $\mathbf{Q}_\pm = (0.5, 0.5 \pm 0.23, 0.5)$. With the understanding that the two $\mathbf{Q}_\pm$ positions are symmetric around (0.5, 0.5, 0.5), they are physically equivalent. The data suggest that the signal amplitude is slightly larger at $\mathbf{Q}_-$, consistent with the expectation that the form factor decreases with increasing $|\mathbf{Q}|$ for magnetic scattering. The data also show that there is no intensity enhancement below $T_\mathrm{c}$ at (0.5, 1.0, 0.5) which is equivalent to (0.5, 0, 0.5), where the strongest increase of magnetic signals below $T_\mathrm{c}$ is commonly observed in FeSCs with hole pockets at the $\Gamma$ point \cite{DaiRevModPhys2015}. We emphasize that the total magnetic signals at $\mathbf{Q}_\pm$ may be substantially greater than the peak amplitudes displayed in Fig.~\ref{Fig2}(b), which only correspond to the part of signals that changes below $T_\mathrm{c}$. The total magnetic signal is difficult to estimate from our data due to the presence of strongly $\mathbf{Q}$-dependent phonon scattering at nearby energies [Fig.~\ref{Fig3}(a)], which explains the strong variation of intensity versus $\mathbf{Q}$ in Fig.~\ref{Fig2}(a).

\begin{figure}
\includegraphics[width=3.375in]{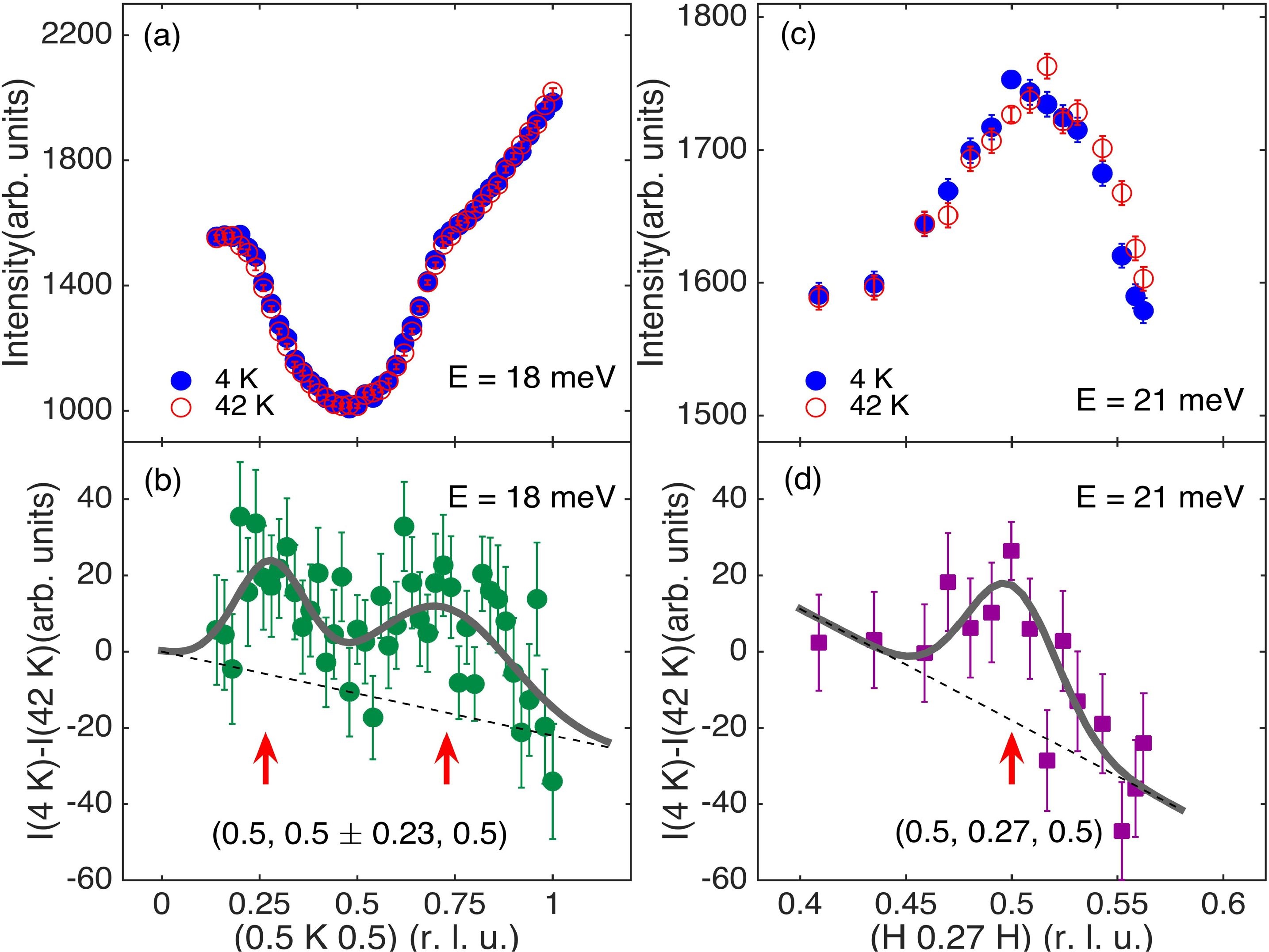}
\caption{\label{Fig2}
(a) $\mathbf{Q}$ scans along the (0.5, $K$, 0.5) direction at a fixed energy transfer of 18 meV and temperatures of 4 K and 42 K. (b) Intensity difference between the two temperatures.  (c) $\mathbf{Q}$ scans at 21 meV in a direction perpendicular to that in (a). (d) The intensity difference between the two temperatures in (c). Arrows indicate the fitted peak centers.
}
\label{Fig2}
\end{figure}

In order to pin down the $\mathbf{Q}$ position that has the strongest ($T$-dependent) magnetic signal, we have performed $\mathbf{Q}$ scans along the perpendicular direction ($H$, 0.27, $H$) going through $\mathbf{Q}_-$, and the data [Fig.~\ref{Fig2}(c-d)] confirm that the magnetic signals are centered on the high-symmetry line (0.5, $K$, 0.5). Therefore, the distribution of magnetic signals in the $\mathbf{Q}$ space is overall very similar to that in $A_x$Fe$_{2-y}$Se$_2$ ($A=$ Rb, Cs, K) \cite{ParkPhysRevLett2011,TaylarPhysRevB2012,FriemelEPL2012}. A similar interpretation, that the characteristic $\mathbf{Q}_\pm$ follow from a FS-nesting picture with sign-reversal pairing \cite{MaierTAPhysRevB2011,KhodasPRL2012}, also applies here.

\begin{figure}
\includegraphics[width=3.375in]{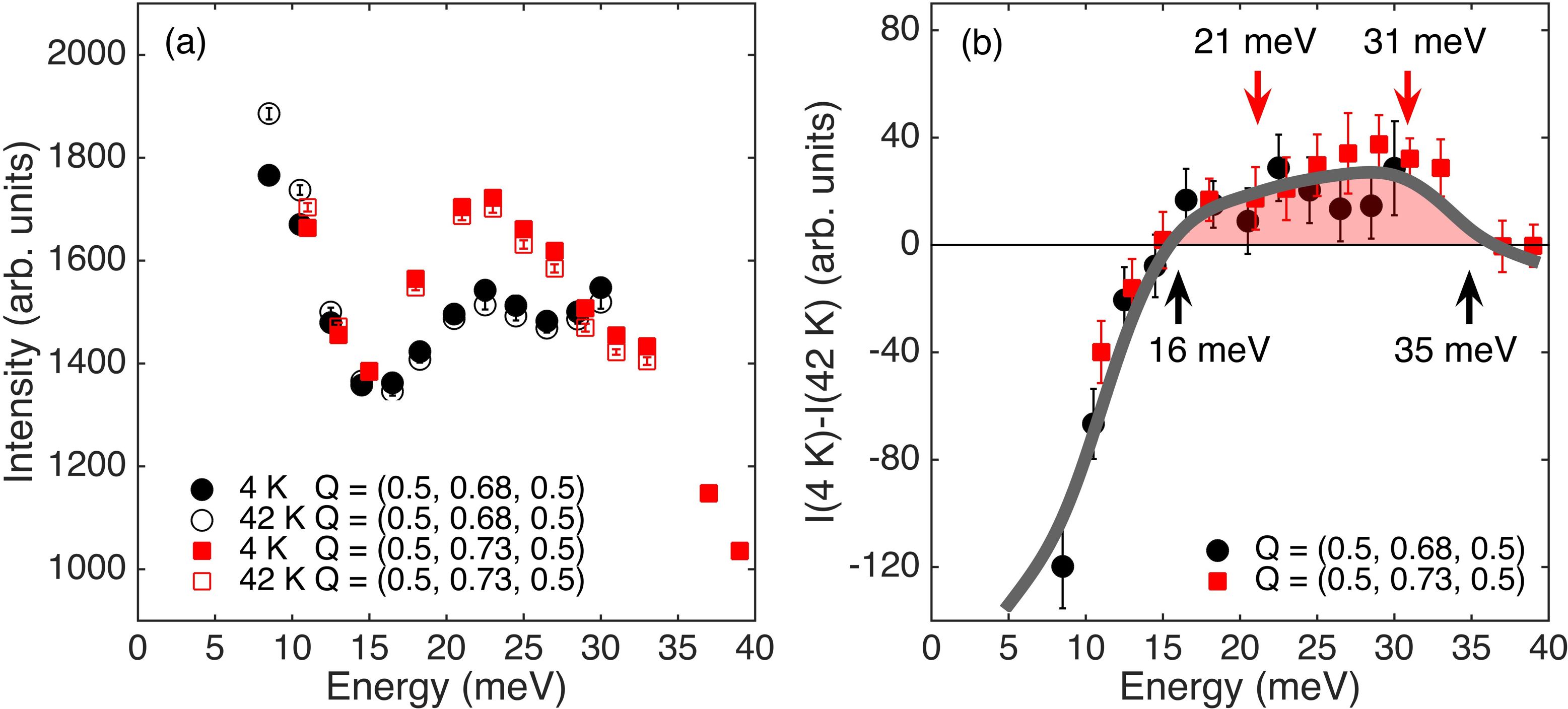}
\caption{\label{Fig3}
(a)  Energy scans at fixed $\mathbf{Q}$ positions (0.5, 0.68, 0.5) and (0.5, 0.73, 0.5) in the superconducting state ($T = 4$ K) and normal state ($T = 42$ K). (b) Intensity difference between the two temperatures.
}
\label{Fig3}
\end{figure}

Figure~\ref{Fig3} displays energy scans performed at $\mathbf{Q}$ = (0.5, 0.68, 0.5) and (0.5, 0.73, 0.5) over an extensive energy range, and again we use the intensity difference between 4 K and 42 K to reveal the magnetic signals. We consider both of these $\mathbf{Q}$ positions to be sufficiently close to $\mathbf{Q}_+$ given the large momentum widths of the signals in Fig.~\ref{Fig2}(b). Similar measurements cannot be performed near $\mathbf{Q}_-$ due to neutron kinematic constraints in the scattering process. Surprisingly, it turns out that the energies that we have chosen (18 and 21 meV) for the measurements in Fig.~\ref{Fig2} belong to a rather broad distribution of intensity enhancement below $T_\mathrm{c}$, ranging from 16 meV to about 35 meV. In fact, the globally greatest net intensity increase occurs at about 30 meV. Below 16 meV, a strong decrease of intensity is observed upon cooling from 42 K to 4 K, due to both  opening of the superconducting gap and reduction in thermally-activated phonon scattering.

Motivated by the broad energy distribution, we have performed detailed measurements of the $T$ dependence of the scattering signal, at both 21 meV and 31 meV. The results are displayed in Fig.~\ref{Fig4}. For 21 meV the measurement was performed at $\mathbf{Q}_-$, where the kinematic constraints can still be satisfied and the signal is stronger, whereas for 31 meV the measurement could only be performed at $\mathbf{Q}_+$. The difference in $\mathbf{Q}$ explains the seemingly different relative intensities at these two energies comparing Fig.~\ref{Fig4} and Fig.~\ref{Fig3}(b). At 21 meV, the magnetic signal exhibits an order-parameter-like increase below $T_\mathrm{c}$, and is therefore consistent with being a spin resonance \cite{YuNatPhys2009,DaiRevModPhys2015}. In contrast, the signal at 31 meV exhibits a continuous increase towards the lowest temperature with only a weak (if any) anomaly at $T_\mathrm{c}$.

\begin{figure}
\includegraphics[width=3.375in]{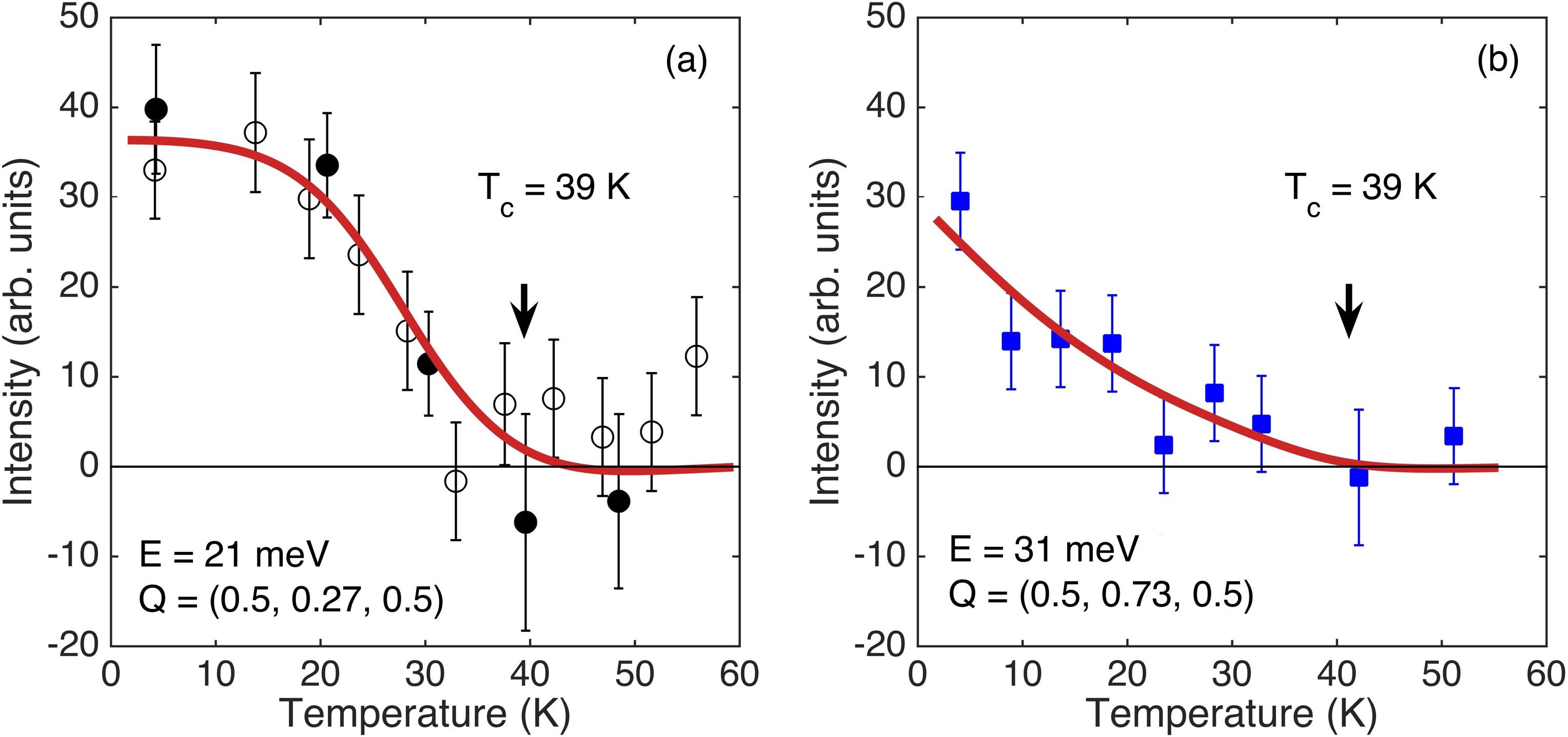}
\caption{\label{Fig4}
(a) $T$ dependence of INS signal at $\mathbf{Q} =$ (0.5, 0.27, 0.5) and 21 meV. (b) $T$ dependence of INS signal at $\mathbf{Q} =$ (0.5, 0.73, 0.5) and 31 meV. Averaged intensities just above $T_\mathrm{c}$ are set to zero. Solid lines are guide to the eye.
}
\label{Fig4}
\end{figure}

Very recently and after the completion of our experiments, there were two reports of INS studies of (Li$_{0.8}$Fe$_{0.2}$)ODFeSe using powder \cite{DaviesPhysRevB2016} and single-crystal \cite{Pan2016} samples. Here we compare our data to the previously reported ones, as there are several quantitative differences at first glance: (1) The in-plane positions of $\mathbf{Q}_\pm$ were estimated to be $(0.5, 0.5\pm0.17)$ in Ref.~\onlinecite{DaviesPhysRevB2016} and $(0.5, 0.5\pm0.18)$ in Ref.~\onlinecite{Pan2016}, whereas we find $(0.5, 0.5\pm0.23)$. (2) The distribution of magnetic intensities in energy appears quite different. The spin resonance was reported to be centered at $\approx23$ meV \cite{DaviesPhysRevB2016} and 21 meV \cite{Pan2016}, with relatively weak and nearly no $T$-dependent intensities above 28 meV $\approx2\Delta_\mathrm{SC}$, respectively, whereas here we still find a substantial increase of magnetic scattering below $T_\mathrm{c}$ at 31 meV despite the distinct temperature dependence in Fig.~\ref{Fig4}. (3) The momentum width of the resonance was previously found to be larger in the transverse direction with respect to the displacement away from (0.5, 0.5) \cite{Pan2016}. In our data, the width appears to be larger in the longitudinal direction (Fig.~\ref{Fig2}), although the statistical accuracy is still too limited for us to make a firm conclusion here.

An important difference between our measurements and the previous single-crystal study \cite{Pan2016} is the out-of-plane momentum transfer, which was chosen here to be 0.5 r.l.u. and zero in the previous triple-axis INS experiments. It is possible that the energy and in-plane momentum distribution of magnetic signals is sensitive to the choice of the out-of-plane momentum transfer, because of FS warping \cite{ParkPRB2010,FriemelPRB2012} and/or variation of $2\Delta_\mathrm{SC}$ with $k_z$ \cite{ZhangNatPhys2012}. To our knowledge, no similar investigations have been reported so far for (Li$_{0.8}$Fe$_{0.2}$)OHFeSe. Indeed, according to the time-of-flight INS data displayed in Figs.~2 and 3 of Ref.~\onlinecite{Pan2016}, which integrate over a wide range of out-of-plane momentum transfer, the low-energy magnetic signals are centered at in-plane $\mathbf{Q}$ positions very close to our result. Moreover, in the previous powder study \cite{DaviesPhysRevB2016} where magnetic signals integrated over a broad range in $|\mathbf{Q}|$ were used to construct the energy distribution, a noticeable amount of intensity increase below $T_\mathrm{c}$ was present above $2\Delta_\mathrm{SC}$. We therefore conclude that the above differences are primarily related to the out-of-plane momentum transfer, although additional variations related to the samples, such as the precise electron doping, may also play a role.

The overall phenomenon in Fig.~\ref{Fig3} is somewhat similar to the ``even-parity'' spin resonance in heavily hole-doped high-$T_\mathrm{c}$ cuprates, in which case the feature becomes very broad as its energy approaches $2\Delta_\mathrm{SC}$ \cite{PailhesPRL2003,CapognaPRB2007}, whereas the ``odd-parity'' counterpart at lower energy is considerably sharper. Although the presence of two resonant modes has also been reported in the pnictides \cite{ZhangPhysRevLett2013,SteffensPhysRevLett2013}, we believe that it is unrelated to our observation, since the signals at 21 and 31 meV have different $T$ dependence, and because 31 meV is clearly \textit{above} $2\Delta_\mathrm{SC}$ \cite{ZhaoLNatureCommun2015,NiuPhysRevB2015,DuNatureCommun2016}. The fact that the response above $2\Delta_\mathrm{SC}$ is smoothly connected to the resonance at 21 meV \cite{DaviesPhysRevB2016,Pan2016} suggests that the superconductivity in (Li$_{0.8}$Fe$_{0.2}$)OHFeSe might be supported by two types of pairing interactions. The first type, presumably related to spin fluctuations, is in favor of pairing in a sign-reversed fashion which leads to the formation of a spin resonance below $2\Delta_\mathrm{SC}$; the second type, possibly related to electron-phonon interactions and/or orbital fluctuations, is in favor of plain $s$-wave pairing, which simply leads to a pile-up of magnetic spectral weights just above $2\Delta_\mathrm{SC}$ \cite{OnariPhysRevB2010} in the superconducting state. The apparent dependence of the signals above $2\Delta_\mathrm{SC}$ on the out-of-plane momentum transfer, as discussed above, suggests that the second type of pairing interactions might be further selective to the $k_z$ quantum number of the quasiparticles.

Finally, our results are likely related to the recent observation of similar phenomena in sulfur-doped K$_x$Fe$_{2-y}$Se$_2$ \cite{WangPhysRevLett2016}, which were interpreted as evidence for a transition from sign-reversed to sign-preserved Cooper pairing. The transition was observed upon a simultaneous suppression of $2\Delta_\mathrm{SC}$ and $T_\mathrm{c}$ by sulfur doping, which does not seem to affect the characteristic energies of the spin fluctuations. It is therefore plausible that the second type of pairing interactions in sulfur-doped K$_x$Fe$_{2-y}$Se$_2$ are capable of supporting superconductivity with $T_\mathrm{c}$ up to $\approx25$ K \cite{WangPhysRevLett2016}. In comparison to that, (Li$_{0.8}$Fe$_{0.2}$)OHFeSe can be viewed as FeSe sheets intercalated with molecules containing much lighter atoms. Hence, if the second type of pairing interactions are associated with phonons in structural units next to the FeSe sheets, such as in the substrate of FeSe/STO \cite{LeeNature2014,LiSciBull2016}, the interactions will be able to support superconductivity in (Li$_{0.8}$Fe$_{0.2}$)OHFeSe up to higher temperatures than in K$_x$Fe$_{2-y}$Se$_2$. This is consistent with our observation of magnetic signals above $2\Delta_\mathrm{SC}$ already in a $T_\mathrm{c}=39$ K sample, as well as with the experimental evidence for predominant plain $s$-wave pairing \cite{FanNatPhys2015} in FeSe/STO with even higher $T_\mathrm{c}$.

\begin{acknowledgments}
We wish to thank Jitae Park and Fa Wang for discussions. This work is supported by the National Natural Science Foundation of China (Grants No. 11374024 and No. 11522429) and Ministry of Science and Technology of China (Grants No. 2015CB921302 and No. 2013CB921903).
\end{acknowledgments}

\preprint{Preprint}

\bibliography{Reference}

\end{document}